\documentstyle{article}

\title{
An Entire Spectral Determinant for Semiclassical Quantization
}
\author{G\'abor Vattay \\
{\em \small Niels Bohr Institute, Blegdamsvej 17, DK-2100 Copenhagen \O,
Denmark} \\
and \\
{\em \small E\"otv\"os University, M\'uzeum krt. 6 - 8, H-1088 Budapest,
Hungary}}

\date{ }

\begin{document}
\maketitle
\begin{abstract}
We show that the eigenvalues of the first order partial
differential equation derived by quasi-classical approximation of
the Schr\"odinger equation can be computed from the trace of
a classical operator. The derived trace formula is different
from the Gutzwiller trace formula. The Fredholm determinant
of the new operator is an entire function of the complex energy
plane in contrast to the semi-classical spectral determinant
derived from the Gutzwiller trace formula.
\end{abstract}
%\vspace{10cm}
%\pagebreak

\baselineskip.55cm

\section{Introduction}
The Gutzwiller trace formula\cite{Gut} is the most important tool for
semiclassical quantization of non-integrable systems.
The trace formula,
in general, does not converge close to the semiclassical energy
eigenvalues or resonances
where we would like to use it. The spectral determinant,
derived by Voros\cite{Voros} from the trace formula, has much better
mathematical properties. However it also does not converge on the whole
complex energy plane\cite{EckRuss} and cannot be used to
determine all the complex resonances of a system. One possible
solution of the convergence problem is to introduce smeared trace
formulas or spectral determinants. These methods limit the accuracy
of the resonances and allows only qualitative comparison of
the exact quantum and the semiclassical spectra.
Trace formulas and spectral determinants can be introduced for
classical systems\cite{Ruelle} including thermodynamic
formalism\cite{thermo} and their analytical properties
can be studied. The convergence problems of these formulas
and the semiclassical convergence problems have the same origin
\cite{CRRV}.
For a large class of classical systems with `nice' mathematical properties
the recent theorem of H. H. Rugh\cite{Rugh} assures the convergence
of classical spectral (Fredholm) determinants for a large class of
classical systems with `nice' mathematical properties.
This theory can
be extended to semiclassical trace formulas\cite{CV}, and the
semiclassical resonances can be computed without convergence problems.
In addition, the numerical error of the semiclassical resonances
decreases super-exponentially $\sim\exp(-n^\beta)$ ($\beta>1$)
as a function of the maximal topological length $n$
of  periodic orbits.

The theory in Ref.\cite{CV} has been developed on an abstract mathematical
ground. Here, the connection of this mathematical description and
the {\em quasi-classical} approximation, introduced by Maslov\cite{Maslov},
is discussed. The term quasi-classical is more appropriate than
semiclassical since
the Maslov type description leads to a pure classical
Perron-Frobenius operator in a natural way.
The Fredholm determinant of this operator can serve as a well defined
and mathematically clean starting point of the semiclassical approximation
of the spectra.
One additional advantage of this description
is that the wave function evolves along one single classical trajectory
and we do not have to compute sums over increasing numbers of classical
trajectories as in computations
involving Van Vleck formula\cite{Heller}.

\section{The quasi-classical approximation}
Following mostly Ref.\cite{Arnold}, we give  a summary of the
quasi-classical approximation, which was worked out by Maslov\cite{Maslov}
in this form.
This approximation was used earlier by Liouville, Green, Stokes,
Rayleigh and others.

The Schr\"o\-din\-ger equation for a single particle of unit mass
in a potential
$U$ is
\begin{equation}
i\hbar \frac{\partial \psi}{\partial t}= -\frac{\hbar^2}{2}\Delta \psi +
U(q)\psi, \label{sch}
\end{equation}
where $\Delta$ is the Laplace operator and
$\psi(q,t)$ is the wave function.
The $\hbar$ is the small parameter of the problem, and the
quasi-classical approximation gives the solution of this equation
for $\hbar\rightarrow 0$. This asymptotic solution
is of short wave form
\begin{equation}
\psi(q,t)=\varphi(q,t) e^{iS(q,t)/\hbar},\label{sh_wave}
\end{equation}
where the amplitude $\varphi(q,t)$ and the phase
$S(q,t)$ are smooth real functions on some
bounded region of the configuration space.
Substituting (\ref{sh_wave}) in the Schr\"o\-din\-ger equation,
we can derive partial differential equations (PDE) for
the amplitude and the phase with initial conditions
$\varphi(q,0)=\varphi_0(q)$ and $S(q,0)=S_0(q)$ respectively.
In the limit $\hbar\rightarrow 0$ omission of the non-leading
terms proportional to $\hbar^2$ leads to the following quasi-classical
partial differential equations
\begin{eqnarray}
\frac{\partial \varrho}{\partial t}+ div \left( \varrho \nabla
S\right)&=&0, \label{cont}\\
\frac{\partial S}{\partial t} +H(q,\nabla S) &=&0,\label{ham_jac}
\end{eqnarray}
where $\varrho(q,t)=\varphi^2(q,t)$ and $H(q,p)$ is the Hamilton function.
Equation
(\ref{ham_jac}), which is the Hamilton-Jacobi equation, is
a first order partial differential equation whose solution
corresponding to an initial $S(q,0)=S_0(q)$, can be given independently
from (\ref{cont}). However, the continuity equation
(\ref{cont}) can be solved only if the solution $S(q,t)$ of
(\ref{ham_jac}) is known.
In other words, the Hamilton-Jacobi equation is autonomous
while the continuity equation is driven by the solution
$S(q,t)$ of (\ref{ham_jac}).

We know from the theory of first order PDE's\cite{Arnold2}
that their solutions lead to {\em ordinary} differential equations (ODE).
As is well known, the Hamilton-Jacobi equation
leads to the Hamilton differential equations
\begin{eqnarray}
\dot{q}&=&\;\;\frac{\partial H(q,p)}{\partial p}, \nonumber \\
\dot{p}&=&-\frac{\partial H(q,p)}{\partial q},
\end{eqnarray}
where the new variable
$$p=\nabla S(q,t)$$
has been introduced.
In the PDE description we evolve the whole function $S(q,t)$ and
compute its gradient at a given point $q_0$.
Computation of the gradient requires information
about the behavior of the function $S(q,t)$ in the vicinity
of $q_0$ and can not be recovered from the value of $S(q_0,t)$ alone.
In the ODE description we evolve both $q$ and $p=\nabla S(q,t)$.
Therefore at a given time, we do not have
to compute the gradient from $S(q,t)$ since the evolution is {\em local}
in the $(q,p)$ space. From the ODE description we can
reconstruct the whole solution of the PDE as
\begin{equation}
S(q',t)=S(q,0)+\int_0^{t}L(q(\tau),\dot{q}(\tau))d\tau,\label{act}
\end{equation}
where we have to integrate the Lagrange function along the
phase space trajectory with
$$q'=q(t), q=q(0), \nabla S(q',t)=p(t),
\nabla S(q,0)=p(0).$$
For a generic $S(q,0)$ only one such trajectory exists.

The  `local'  solution of the
continuity equation (\ref{cont}) is also straightforward.
At a given starting point $q_0$ the momentum is given
by $p_0=\nabla S_0(q_0)$, and the amplitude of the
wave function is $\varphi_0(q)$. After time $t$
the coordinate $q_0$ evolves to $q(t)$ and $p_0$ to $p(t)$
according to the full Hamiltonian flow.
The new amplitude can be derived from the probability conservation
as follows: Take an infinitesimal initial $d$
dimensional directed volume $V(q_0)$
around $q_0$ in the configuration space. The points of
this volume evolve to
the infinitesimal directed volume $V(q(t))$ around $q(t)$
according to the Hamiltonian flow. The value of the
amplitude changes according to
\begin{equation}
\varphi(q(t))=\pm
\left(\frac{V(q(t))}{V(q(0))}\right)^{-1/2}\varphi_0(q_0),
\end{equation}
where the sign $\pm 1$ reflects the ambiguity of the
transformation from the density $\varrho$ to $\varphi$.
Careful analysis shows that the minus sign refers
to the case where the orientation of the final volume
is the opposite the initial one.
The ratio of volumes is independent of the way we specify the
initial infinitesimal volume.
To understand which quantities determine the volume ratio,
we specify an initial directed parallelepiped around $q_0$
with edges given by $d$ independent
infinitesimal vectors
$\delta {\bf q}_1,\delta {\bf q}_2,\ldots \delta {\bf q}_d$.
After time $t$ these vectors are transformed into
$\delta {{\bf q}'}_1,\delta {{\bf q}'}_2,\ldots \delta {{\bf q}'}_d$
by the flow. The initial point $(q_0,p_0)$ and the set of
initial vectors do not specify uniquely the image vectors.
This is a consequence of the fact that the infinitesimal vectors are
transformed by
the full Jacobian matrix of the Hamilton flow, and
we have not specified the infinitesimal momentum vectors around
$p_0$ corresponding to these vectors. The initial
function $S_0(q)$ determines a set
$\delta {{\bf p}}_1,\delta {{\bf p}}_2,\ldots
\delta {{\bf p}}_d$ of vectors through the second
derivative matrix:
\begin{equation}
\delta {{\bf p}}={\bf M}\delta {{\bf q}},\;\;
{\bf M}_{ji}=\frac{\partial^2 S_0(q)}{\partial q_j \partial
q_i},\label{secdel}
\end{equation}
which we shall call the {\em curvature matrix}.
The initial curvature matrix ${\bf M}_0$ is an important
initial condition, and
we are not able to compute the image of the volume without it.
The vector $(\delta {\bf q}_i,\delta {\bf p}_i)$
is transformed by the Jacobi matrix
\begin{eqnarray}
\delta {{\bf q}'}_i&=&{\bf J}_{qq}\delta {\bf q}_i+{\bf J}_{qp}\delta
{\bf p}_i, \\
\delta {{\bf p}'}_i&=&{\bf J}_{pq}\delta {\bf q}_i+{\bf J}_{pp}\delta
{\bf p}_i, \\
\end{eqnarray}
where the subscripts $q$ and $p$ denote the corresponding
$[d\times d]$ submatrices of the full $[2d\times 2d]$ Jacobian.
The Jacobian is determined by the initial condition
$(q_0,p_0)$ and can be computed as a time ordered integral
along the phase space trajectory
\begin{equation}
{\bf J}(q,p,t)=
\mbox{T} \exp\left\{ \int_{0}^{t} d\tau {\bf D}^2 H(q(\tau),p(\tau)) \right\},
\end{equation}
where  ${\bf D}^2 H(q,p)$ denotes the second symplectic
derivative matrix of the Hamiltonian and $\mbox{T}$ is the time
ordering operator.
The curvature matrix of the function $S(q,t)$ around
$q(t)$ defines a (\ref{secdel}) type relation between
the infinitesimal vectors
\begin{equation}
\delta {{\bf p}'}={{\bf M}'}\delta {{\bf q}'},\;\;
{{\bf M}'}_{ji}=\frac{\partial^2 S(q',t)}{\partial q_j \partial
q_i}.\label{secdel2}
\end{equation}
Using  (\ref{secdel}) and  (\ref{secdel2}), we can eliminate
the vectors $\delta {{\bf p}}_i$ and $\delta {{\bf p}'}_i$
and can get relations between the initial and final
infinitesimal configuration vectors and the curvature matrices
\begin{eqnarray}
\delta {{\bf q}'}&=&({\bf J}_{qq}+{\bf J}_{qp}{\bf M})\delta {\bf q},
\label{qprime}\\
{\bf M}'&=&({\bf J}_{pq} + {\bf J}_{pp}{\bf M})({\bf J}_{qq} + {\bf
J}_{qp}{\bf M})^{-1}.\label{mprime}
\end{eqnarray}
{}From the first relation (\ref{qprime}) we can read off the volume
ratio:
\begin{equation}
\frac{V(q')}{V(q_0)}=
\det({\bf J}_{qq}+{\bf J}_{qp}{\bf M}).
\end{equation}
The second relation (\ref{mprime}) is a recursion relation for
${\bf M}$ and can be considered as the matrix generalization of the
usual fractional rational transformation.
{}From (\ref{mprime}) we can derive a differential equation for ${\bf
M}(t)$, if we substitute the elements of the infinitesimal Jacobi matrix.
This differential equation
\begin{equation}
\dot{{\bf M}}=-\left( \frac{\partial^2 H}{\partial q\partial q}
+{\bf M}\frac{\partial^2 H}{\partial p\partial q}+\frac{\partial^2 H}{\partial
q\partial p}{\bf M}+{\bf M}\frac{\partial^2 H}{\partial p\partial p}{\bf M}
\right),\label{matric}
\end{equation}
is a driven one since the second partial derivatives of the
Hamilton function should be computed along the phase space trajectory.
If we solve this differential equation along the phase space trajectory
the volume ratio can be
written as a time ordered integral along the phase space and ${\bf
M}(t)$ trajectory
\begin{equation}
\frac{V(q')}{V(q_0)}=\mbox{T} \exp\left\{ \int_{0}^{t} \mbox{Tr}\left[
\frac{\partial^2 H}{\partial p\partial q}+\frac{\partial^2
H}{\partial p\partial p}{\bf M}
\right]d\tau\right\}
\end{equation}
The square root of the volume ratio is also a time ordered
integral:
\begin{equation}
\left(\frac{V(q')}{V(q_0)}\right)^{-1/2}=\mbox{T} \exp\left\{-\frac{1}{2}
\int_{0}^{t} \mbox{Tr}\left[
\frac{\partial^2 H}{\partial p\partial q}+\frac{\partial^2
H}{\partial p\partial p}{\bf M}
\right]d\tau\right\} .\label{xx}
\end{equation}
The computation of this expression requires some care when the
solution of the differential equation (\ref{matric}) is
singular. Close to a singularity, where $${\bf M}(t\rightarrow
t^c)=\infty,$$
we can neglect the non-leading terms from (\ref{matric}) and
use the solution of
\begin{equation}
\dot{{\bf M}}=-{\bf M}\frac{\partial^2 H}{\partial p\partial p}{\bf M}
\label{singsing}.
\end{equation}
The second derivative matrix can be decomposed into combinations of dyads
and their eigenvalues in the usual way
\begin{equation}
\frac{\partial^2 H}{\partial p\partial p}
=\sum_{i=1}^{d}\lambda_i {\bf D}_i, \;\; {\bf D}_i {\bf
D}_j=\delta_{ij}{\bf D}_j.
\end{equation}
The solution close to the singularity can be a linear combination
of some of these dyads corresponding to singular directions $l$:
\begin{equation}
{\bf M}(t)=\frac{1}{t-t^c}\sum_{l=1}^{R}
\frac{1}{\lambda_l}{\bf D}_l,
\end{equation}
where $R$ is the number of singular directions.
The time ordered integral close to the singularity is dominated by
\begin{equation}
\left(\frac{V(q(t^c_{+0}))}{V(q(t^c_{-0}))}\right)^{-1/2}=
\exp\left(-\frac{1}{2}\int_{t_c-0}^{t_c+0}\frac{R}{\tau-t_c}d\tau\right).
\end{equation}
This integral can be computed by adding infinitesimal
imaginary value $i\epsilon$
to $t^c$ and taking the $\epsilon\rightarrow 0$ limit
\begin{equation}
\left(\frac{V(q(t^c_{+0}))}{V(q(t^c_{-0}))}\right)^{-1/2}=\exp(i\pi(R/2)).
\end{equation}
Between two singular points the time ordered integral is positive
and gives the absolute value of the volume ratio. Notice that
$R$ counts the number of rank reductions of the matrix ${\bf M}$
along the classical path, and it is also
a function on the initial condition ${\bf M}_0$.

%We can notice, that these equations are simpler
%if we take into consideration the
%energy conservation and redo the derivation in a coordinate system
%orthogonal to the direction of the flow. This way we can eliminate one
%spatial dimension.
%For example, in two-dimensional billiard systems
%the ${\bf M}$ $2 \times 2$ matrix reduces to one number, the so-called
%Sinai-Bunimovich curvature, ${\bf M}=\kappa$, and (\ref{matric}) reduces
%to the Riccati equation
%\begin{equation}
%\dot{\kappa}=-\kappa^2 - K(t)
%\end{equation}
%where $K(t)$ is the Riemann curvature of the space along a trajectory
%at time $t$.

Now we have everything needed to describe the time evolution
of a quasi-classical wave function. The wave function
at time $t$ is now
\begin{equation}
\psi(q',t)=\varphi(q',t)
e^{iS(q',t)/\hbar}=\pm\left(\frac{V(q')}{V(q_0)}\right)^{-1/2}
e^{i\int_0^{t}Ld\tau/\hbar}
\varphi_0(q_0)e^{iS_0(q_0)/\hbar} \label{VV}
\end{equation}
where $q_0$ is the starting point of a classical trajectory with
initial momentum $\nabla S_0(q_0)$ which ends up in $q$ after time
$t$ with momentum  $\nabla S(q,t)$, and the volume ratio is
determined by the curvature matrix ${\bf M}=\partial_i \partial_j
S(q,0)$.

\section{Time evolution \`a la Maslov}

At this point it is possible to express the volume ratio and
the momentum with the second and first derivatives of the
minimal action between $q'$ and $q$. In this way we recover
the usual Van Vleck propagator\cite{Littlejohn}.
The spirit of equation (\ref{VV}) is that
the wave amplitude $\varphi$ at time $t$ and at coordinate
$q'$ is determined by  the amplitude at $t=0$ at coordinate $q$.
In calculations involving the Van Vleck operator kernel
this nice property is lost, and we have to compute lots of
trajectories to compute the volume ratio and we have to know
the whole initial wave function too.
However we have a better option.
We can keep track of the variables $p$ and ${\bf M}$ along only
one trajectory and compute (\ref{act}) and the volume ratio
(\ref{xx}). This means that the evolution takes place on the
extended $(q,p,{\bf M})$ space.
We can introduce classical density  functions $\tilde{\psi}$
defined on this space.
The wave function then corresponds to the special function
\begin{equation}
\tilde{\psi}(q,p,{\bf M},t)=\psi(q,t)\delta(p-\nabla S(q,t))
\delta\left({\bf M}-
\frac{\partial^2 S(q,t)}{\partial q_j \partial q_i}\right).\label{tild}
\end{equation}
The evolution of a general classical density function
on the extended space according to (\ref{VV})
can be rewritten in terms of a classical transfer
operator
\begin{equation}
\tilde{\psi}(q',p',{\bf M}',t)=\int dqdpd{\bf M}{\cal L}(q',p',{\bf
M}',t\mid q,p,{\bf M},0)\tilde{\psi}(q,p,{\bf M},0),
\end{equation}
with the kernel
\begin{equation}
e^{i\pi\nu+\int_0^{t}d\tau \frac{iL}{\hbar}+
\frac{1}{2}\mbox{Tr}\left\{\frac{\partial^2 H}{\partial p\partial
q}+\frac{\partial^2
H}{\partial p\partial p}{\bf M}\right\}}
\delta(q'-q^{t}(q,p))\delta(p'-p^{t}(q,p))
\delta({\bf M}'-{\bf M}^{t}(q,p,{\bf M})),
\label{main_res}
\end{equation}
where $q^{t}(q,p)$, $p^{t}(q,p)$ and ${\bf M}^{t}(q,p,{\bf
M})$ denote the evolution of $q$, $p$ and ${\bf M}$ from the
initial coordinates $q$,$p=\nabla S_0(q)$ and
${\bf M}=\partial_i \partial_j S_0(q)$ during the
time $t$, and $\nu=N+R/2$. The time ordered integrals
should be computed along the full trajectory, and also the number
of rank reductions $R$ and the number of orientation changes $N$.
The Dirac deltas assure that the operator connects
coordinates, which are connected by the classical dynamics,
and give the correct amplitude.
This operator can evolve densities, which are not of the form (\ref{tild}),
and therefore we can expect that only a part of its spectrum has relevance
to semiclassics, but all the semiclassical eigenvalues will be contained
in its spectra.

\section{Entire Fredholm determinants}

The spectral or Fredholm determinant of the operator (\ref{main_res})
can be defined by the identity
\begin{equation}
\det(1-{\cal L})=\exp\left(-\sum_{n=1}^{\infty}\frac{1}{n}\mbox{Tr}{\cal
L}^n\right),
\end{equation}
where the traces of the powers of the operator are
\begin{equation}
\mbox{Tr}{\cal L}^n=\int dqdpd{\bf M} {\cal L}^n(q,p,{\bf M}\mid q,p,{\bf
M}).
\end{equation}
The zeroes of the Fredholm determinant on the complex energy plane
yield the eigenvalues of the operator.
The simplest application of (\ref{main_res}) is to
3-dimensional hyperbolic Hamiltonian flows. In this case
the spectral or Fredholm determinant of the operator (\ref{main_res})
is given by\cite{CV}
\begin{eqnarray}
\det({\bf 1}-{\cal L})\,\,&=&\exp\left(-\sum_{p,r}
        \frac{1}{r \mid \Lambda_p^r \mid}
      \frac{e^{irS_p(E)+i\nu_p\pi}}{(1-1/\Lambda_p^r)^2}
        \Delta_{p,r}
\right)
                \nonumber \\
 \Delta_{p,r} &=&
      \frac{\mid \Lambda_p^r \mid^{1/2} }{1-1/\Lambda_p^{2r}}
      + \frac{ \mid \Lambda_p^r \mid^{-5/2}}{1-1/\Lambda_p^{2r}}
\,,
\label{Ham_FD}
\end{eqnarray}
where $p$ is the index of primitive periodic orbits, $\Lambda_p$ is the
corresponding eigenvalue of the stability matrix, and
$r$ is the repetition number.
The practical advantage of (\ref{Ham_FD}) over
the more familiar Gutzwiller-Voros and Ruelle type zeta functions
was demonstrated by detailed numerical studies\cite{CRRV}
of the related quantum Fredholm determinant\cite{CR92}.
It can be shown\cite{vattay} that the Fredholm determinant
obtained by keeping only one of the terms in the sum in (\ref{Ham_FD})
is an entire function in the whole energy plane.
This enables us to show that the Gutzwiller-Voros
spectral determinant for Axiom A flows is meromorphic in
the complex $E$ plane, as it can be written as a ratio of entire
functions\cite{CV}.  The non-physical eigenvalues of (\ref{main_res})
can be removed.

\section{Conclusions}
In conclusion, we have constructed a classical evolution
operator for semi-classical quantization based on
Maslov's quasi-classical quantization method, and derived a
new determinant for periodic orbit quantization of chaotic
dynamical systems.
The main virtue of the determinant (\ref{Ham_FD})
is that the theorem of H.H.~Rugh\cite{Rugh}, applicable to
multiplicative
evolution operators such as (\ref{main_res}), implies that this
determinant should be entire for Axiom A flows,
{\em i.e.} free of poles in the entire complex energy
plane.
Our numerical tests of the three disk system
also support the above claims\cite{vattay}.

\section{Acknowledgments}
The author thanks useful comments from
P. Cvitanovi\'c, P. E. Rosenqvist, S. Creagh, P. Dimon and A. Wirzba, and
is grateful to the Sz\'echenyi Foundation, the George Soros
Foundation, OTKA F4286 and the organizers for support,
and to the Center for Chaos
and Turbulence Studies, Niels Bohr Institute for hospitality.


\begin{thebibliography}{99}
\bibitem{Gut} M. C. Gutzwiller, {\em Chaos in Classical Mechanics}
(Springer, New York 1990)
\bibitem{Voros} A. Voros, J. Phys. {\bf A 21}, 685 (1988)
\bibitem{EckRuss} B. Eckhard and G. Russberg, Phys. Rev. {\bf E 47},
1578 (1993)
\bibitem{Ruelle} D. Ruelle, {\em Statistical Mechanics,
Thermodynamical Formalizm} (Addison-Wesley, Reading MA, 1987)
\bibitem{thermo} P. Sz\'epfalusy, T. T\'el, A. Csord\'as and
Z. Kov\'acs, Phys. Rev. {\bf A 36}, 3525 (1987)
\bibitem{Rugh} H. H. Rugh, Nonlinearity {\bf 5}, 1237 (1992) and
H. H. Rugh, {\em Thesis} (1993)
\bibitem{CRRV} P. Cvitanovi\'c, P. E. Rosenqvist, H. H. Rugh
and G. Vattay, {\em Scattering Theory - special issue}, CHAOS (1993)
\bibitem{CV} P. Cvitanovi\'c, G. Vattay, To be published.
\bibitem{Arnold} V. I. Arnold, {\em Mathematical Methods of Classical
Mechanics}, Graduate Texts in Mathematics 60, Springer-Verlag, New York
1987.
\bibitem{Maslov} V. P. Maslov and M. V. Fedoriuk,  {\em Semi-Classical
Approximation in Quantum Mechanics} (Redel, Boston 1981)
\bibitem{Heller} E. J. Heller, S. Tomsovic and A. Sep\'ulveda
CHAOS {\bf 2}, {\em Periodic Orbit Theory - special issue}, 105, 1992
\bibitem{Arnold2} V. I. Arnold, {\em Geometrical Methods in the
Theory of Ordinary Differential Equations}, SpringerVerlag, New York
1983.
\bibitem{Littlejohn} R. G. Littlejohn, Preprint 1991.
\bibitem{CR92} P. Cvitanovi\'c and P.E.~Rosenqvist,
{\em Conference Proceedings}, {\bf 41}, ed.: G. F. Dell'Antonio,
S. Fantoni and V. R. Manfredi (Italian Physical Society, Bologna 1993)
\bibitem{vattay} G. Vattay, To be published.
\end{thebibliography}
\end{document}